\documentclass{PoS}

\usepackage{amsmath}
\usepackage{amsfonts}
\usepackage{amssymb}
\definecolor{rossos}{cmyk}{0,1,1,0.55}
\definecolor{bluscuro}{rgb}{0.15, 0.2, .85}
\definecolor{bluchiaro}{cmyk}{1,.3,0.,0.1}

\newcommand{\eq}[1]{Eq.~(\ref{#1})}

\newcommand{\nn}{\nonumber}

\newcommand{\be}{\begin{equation}}
\newcommand{\ee}{\end{equation}}
\newcommand{\bea}{\begin{eqnarray}}
\newcommand{\eea}{\end{eqnarray}}
\newcommand{\bc}{\begin{center}}
\newcommand{\ec}{\end{center}}

\def\hf{\hat h}
\def\h{h}

\def\stw{s_{\theta_W}}
\def\ctw{c_{\theta_W}}
\def\ttw{t_{\theta_W}}
\title{BSM Primary Effects: The complete set of predictions from the dimension-6 BSM Lagrangian}

\ShortTitle{BSM Primary Effects}

\author{\speaker{Rick S Gupta}\thanks{I thank the organizers of the conference, especially Dr. M Muehlleitner, for inviting me for this talk. This talk is completely based on Ref.~\cite{Gupta}, which was written in collaboration with A. Pomarol and F. Riva.}\\
        IFAE, Universitat Autonoma de Barcelona, 08193 Bellaterra, Barcelona\\
        E-mail: \email{sagupta@cern.ch}}

\abstract{  We present a physical parameterization of the leading effects beyond
the SM (BSM), that give us, at present, the best way to constrain heavy
new-physics at low-energies. We call these effects that constrain all possible interactions at the dimension 6 level,  BSM Primary effects;
there are 8 primaries related to Higgs physics, 3 related to Triple Gauge Couplings and
7 related to $Z$-pole measurements at LEP. Starting from these
experimentally measurable deformations (and not operators), we
construct the dimension 6 Lagrangian in a bottom up way.   We, thus,
show that other BSM effects are not independent from the primary ones and
we provide the explicit correlations. We also discuss the theoretical
expectation for the size of these BSM primaries in some well-motivated
BSM theories. }

\FullConference{XXII. International Workshop on Deep-Inelastic Scattering and Related Subjects,\\
		28 April - 2 May 2014\\
		Warsaw, Poland}

\begin{document}

\section{Introduction}

The first run of LHC was not successful in finding any sign of new physics. This suggests  the presence of a mass gap between the electroweak scale and the scale of new physics.  A model independent way of studying effects of Beyond Standard Model (BSM) physics, in this scenario, is using effective field theory (EFT) to write the BSM Lagrangian as,
\begin{equation}
{\cal L}_{\rm eff}=\frac{\Lambda^4}{g^2_*}
{\cal L}\left(\frac{D_\mu}{\Lambda}\ ,\  \frac{g_* H}{\Lambda}\ ,\  \frac{g_*  f_{L,R}}{\Lambda^{3/2}}\ ,\ \frac{gF_{\mu\nu}}{\Lambda^2}\right)\simeq {\cal L}_4+{\cal L}_6+\cdots\, .
\label{expansion}
\end{equation}
Here ${\cal L}_4$ is the Standard Model (SM) Lagrangian and  ${\cal L}_6$ the dimension 6 extension.  As the number of possible measurable deformations is larger than the  number of couplings at a given order, one can make predictions relating different measurements; for instance, at the dimension 4 level we have the predictions,  $m_Z=m_W/\cos \theta_W,~   Y_f=\sqrt{2} m_f/v$ etc., where $m_W$ and $m_Z$ are the $W$ and $Z$ masses, $Y_f$ is the yukawa coupling, $m_f$ is the fermion mass and $v=246$ GeV the vacuum expectation value (VEV) of the Higgs field. While all the predictions of the the SM Lagrangian are known, the predictions at the dimension 6 level were recently presented for the first time in Ref.~\cite{Gupta}. In this paper we summarise the main results of Ref.~\cite{Gupta}.

We will not use the usual approach of writing a basis of dimension 6 operators and then studying their impact on measurements~\cite{Elias-Miro:2013mua,Pomarol:2013zra}. Instead we will start from measurable deformations and construct the BSM Lagrangian in a bottom up way.  We will show that, apart from four fermion deformations and a gluon self interaction term, the measurement of 18 deformations can constrain all possible terms in the dimension 6 Lagrangian (ignoring CP or minimal flavour violating (MFV) terms). These deformations are the three $(h \bar{f}_L f_R+h.c.)$ deformations, the custodial preserving Higgs boson coupling to $W$ and $Z$ bosons, $ h~A_{\mu\nu}A^{\mu\nu}$, $ h~A_{\mu\nu}Z^{\mu\nu}$, $ h~G_{\mu\nu}G^{\mu\nu}$, $h^3$, seven $Z_\mu \bar{f}\gamma^\mu f$  interactions and three triple gauge coupling (TGC) interactions. These can be, respectively, measured by Higgs production and decay processes at the LHC, the double Higgs production process, $Z$-pole measurements and TGC measurements in the $ee/pp \to WW$ processes. Here and elsewhere, $G_{\mu \nu}$ is the non-abelian gluon field strength and the electroweak field strengths are defined by  $Z_{\mu \nu}\equiv \hat{Z}_{\mu \nu}-i  g \ctw W^+_{[\mu}W^-_{\nu]} $, $A_{\mu \nu}\equiv \hat{A}_{\mu \nu}-i g  \stw W^+_{[\mu}W^-_{\nu]}$ and $W^{\pm}_{\mu\nu}\equiv\hat{W}^{\pm}_{\mu \nu}\pm i  g W^{\pm}_{[\mu}(\stw A+\ctw Z)_{\nu]} $ with $\hat{V}_{\mu\nu}\equiv\partial_\mu V_\nu-\partial_\nu V_\mu$. All other deformations are correlated to these deformations by accidental symmetries at the dimension 6 level. For example, take the BSM primary deformation $ h~A_{\mu\nu}A^{\mu\nu}$, that is related to the  $h \to \gamma\gamma$ branching ratio. At the dimension 6 level, this deformation cannot be generated alone, and as we will show, it must be accompanied by other deformations,
\bea
\Delta {\cal L}^h_{{\gamma\gamma}}&=& \boldsymbol{\kappa_{\gamma\gamma}}  
\left(\frac{ \h}{v}+\frac{\h^2}{2v^2}\right)
\Bigg[ A_{\mu\nu}A^{\mu\nu}+  Z_{\mu\nu}Z^{\mu\nu}+ 2 W^+_{\mu\nu}W^{-\mu\nu}\Bigg]\, .
\label{gamgam}
\eea
This is because $ h~A_{\mu\nu}A^{\mu\nu}$ alone is not a dimension 6 operator, and the other terms are necessary to build a dimension-6 operator. \eq{gamgam} is an example of a BSM primary effect, that includes all the deformations that are generated in a correlated way along with the BSM primary deformation.  In this work we list all the BSM primary effects corresponding to the BSM primaries mentioned above. We will demand that each BSM primary effect should contribute only to the corresponding BSM primary and none other, so \eq{gamgam} above contributes to $h \to \gamma\gamma$ only and not to any of the other 17 BSM primaries. The dimension 6 Lagrangian is a sum of all these independent BSM primary effects.

These BSM Primary effects can be thought of as a basis of observables where each operator contributes to only one observable (see also~\cite{Masso, Elias}). These BSM primary effects are, in our opinion,   more suitable for describing BSM processes than any other basis of dimension 6 operators. This is because  using these, the amplitude for any BSM process  at the dimension 6 level can be directly expressed as a function of the BSM primary measurements, making   the correlations of such a process with all other measurements explicit.

\section{The Higgs primaries}

The first set of primaries we consider are those related to processes involving the Higgs boson. First, we consider  a class of BSM interactions, which affect only Higgs physics and do not affect electroweak symmetry breaking (EWSB) processes such as $Z$-pole or TGC measurements.  Take for instance the deformation in the SM Lagrangian generated by the shift,
\be
Y_f(\hf)= Y_f+\delta Y_f\, {\hf^2}/{v^2}+\cdots
\ee
 where $\hat{h}=v+h$ is the Higgs field.  In the vacuum this just gives a redefinition of the Yukawa couplings in the SM Lagrangian so that the effect of this deformation can be measured only via a process like $h \to ff$, that involves the Higgs excitation $h$. This deformation gives the BSM primary effect,
\bea
\Delta{\cal L}^h_{ff}\!\!&=\!\!&\boldsymbol{\delta g^h_{ff}}\Big(\!\h \bar  f_L f_R+\textrm{h.c.}\!\Big)\!\!
 \left(1+\frac{3\h}{2v}+\frac{\h^2}{2v^2}\right)\,.
 \label{hff}
\eea
In operator language this corresponds to adding $H^\dagger H$ to the dimension 4 SM Yukawa coupling operator, $H$ being the Higgs doublet; \eq{hff} is in fact exactly  this operator. Similarly we can generate  deformations by similar Higgs dependent redefinitions of the  other SM-parameters,
\begin{gather}
e(\hf),\ \stw(\hf),\ g_s(\hf),\ \lambda_h(\hf),\ Z_h(\hf)\, .
\label{generals}
\end{gather}
These generate the BSM primary effects corresponding to the primaries $ h~A_{\mu\nu}A^{\mu\nu}$, $ h~A_{\mu\nu}Z^{\mu\nu}$, $ h~G_{\mu\nu}G^{\mu\nu}$, $h^3$  and  $h~(W^2+Z^2/2\ctw)$, respectively.  Here, $e$ is the electric charge,  $g_s$ is the strong coupling,  $\lambda_h$ is the Higgs quartic coupling in the quartic term $-\lambda_h (H^\dagger H)^2/4$, $Z_h$ is the  coefficient of the Higgs kinetic term and $\stw$ and $\ctw$ are $\sin \theta_W$ and $\cos \theta_W$, respectively.  The explicit results were obtained in Ref~\cite{Gupta} and have been reproduced here in Table 1. Again these BSM primary effects correspond to adding $H^\dagger H$  to operators in the dimension 4 SM Lagrangian.

While, the above class of interactions can definitely be constrained by only Higgs measurements, we have not shown that these are \textit{all} the Higgs primaries. This turns out to be the case, however, because as we will see,  all other terms in the dimension 6 Lagrangian are better constrained by either TGC or $Z$-pole measurements than Higgs physics.

\section{Deviation from universality in the  $Z$-boson couplings }

In the SM the couplings of the $W/Z$ bosons to various particles are related to each other in definite ways due to the gauge coupling universality of the underlying $SU$(2)$\times U$(1) gauge symmetry. Once electroweak symmetry is broken and  the Higgs field gets a VEV the universality of the $W/Z$-boson coupling can be broken by higher dimensional  operators and the coupling to each particle can be individually altered. Including only the first family fermions,  there are seven $Z$-boson couplings (to the three leptons and four quarks), two $W$-boson couplings, the $Z$-coupling to the Higgs field $\hat{h}$ (which can be measured by measuring the $Z$-mass), and the gauge boson self interactions, i.e. the TGCs and the Quartic Gauge Couplings (QGC).  Note that the $W$ coupling to the Higgs field has already been considered when we promoted $Z_h$ to a function of $\hat{h}$ (see $\Delta {\cal L}^h_{VV}$ in Table 1). The analysis in Ref.~\cite{Gupta} reveals two important facts:

\noindent
(1) Not all these couplings can be independently deformed at the dimension 6 level. This is because, there are only seven dimension 6 operators controlling the nine $W/Z$-couplings to fermions so that  the $W$-coupling deformations are not independent of the $Z$-coupling deformations. Thus, we have the following  BSM primary effects for the $W/Z$-coupling to leptons,
\bea
\Delta{\cal L}_{ee}^V&=&\boldsymbol{\delta g^Z_{eR}} \frac{\hf^2}{v^2} Z^\mu \bar e_R\gamma_\mu e_R
+\boldsymbol{\delta g^Z_{eL}}\frac{\hf^2}{v^2} \left[Z^\mu \bar e_L\gamma_\mu e_L-\frac{c_{\theta_W}}{\sqrt{2}}(W^{+\mu}\bar\nu_L\gamma_\mu e_L +\textrm{h.c.}) \right]\nn\\ &&+\boldsymbol{\delta g^Z_{\nu L}}\frac{\hf^2}{v^2} \left[Z^\mu \bar \nu_L\gamma_\mu \nu_L+\frac{c_{\theta_W}}{\sqrt{2}}(W^{+\mu}\bar \nu_L\gamma_\mu e_L +\textrm{h.c.}) \right]
\eea
and a similar expression for the couplings to quarks (see Table 1).
\begin{table}[tc]
\small
\centering
\begin{tabular}{|c|}\hline
Higgs Primaries\\ \hline\\
$\Delta {\cal L}^h_{{\gamma\gamma}}= \boldsymbol{\kappa_{\gamma\gamma}}  
\left(\frac{ \h}{v}+\frac{\h^2}{2v^2}\right)
\Bigg[ A_{\mu\nu}A^{\mu\nu}+  Z_{\mu\nu}Z^{\mu\nu}+ 2 W^+_{\mu\nu}W^{-\mu\nu}\Bigg]\, $\\
$\Delta {\cal L}^h_{{Z\gamma}}=\boldsymbol{\kappa_{Z\gamma}} \left(\frac{\h}{v}+\frac{\h^2}{2v^2}\right)
\Bigg[ \ttw A_{\mu\nu}Z^{\mu\nu} +\frac{c_{2\theta_W}}{2\ctw^2}  Z_{\mu\nu}Z^{\mu\nu}+ W^+_{\mu\nu}W^{-\mu\nu}\Bigg]$\\
$\Delta{\cal L}^h_{{GG}}=\boldsymbol{\kappa_{GG}}\left(\frac{\h}{v}+\frac{\h^2}{2v^2}\right)G^A_{\mu\nu}G^{A\, \mu\nu}\, ,$ \\
$\Delta{\cal L}^h_{ff}=\boldsymbol{\delta g^h_{ff}}\Big(\!\h \bar  f_L f_R+\textrm{h.c.}\!\Big)\!\!
 \left(1+\frac{3\h}{2v}+\frac{\h^2}{2v^2}\right)\,$ \\
$\Delta{\cal L}_{3h}=\boldsymbol{\delta g_{3h}}\, \h^3 \left(1+\frac{3\h}{2v}+\frac{3\h^2}{4v^2}+\frac{\h^3}{8v^3}\right)\ $\\
$ \Delta{\cal L}_{VV}^h=\boldsymbol{\delta g^h_{VV}} \Bigg[
\h
\left( \!W^{+\mu} W^-_\mu+ \frac{Z^\mu Z_\mu}{ 2 \ctw^2}\!\right) +\Delta_{h}\Bigg] $\\ \hline $Z$-pole Primaries\\  \hline\\
$\Delta{\cal L}_{ee}^V=\boldsymbol{\delta g^Z_{eR}} \frac{\hf^2}{v^2} Z^\mu \bar e_R\gamma_\mu e_R
+\boldsymbol{\delta g^Z_{eL}}\frac{\hf^2}{v^2} \left[Z^\mu \bar e_L\gamma_\mu e_L-\frac{c_{\theta_W}}{\sqrt{2}}(W^{+\mu}\bar{\nu_L}\gamma_\mu e_L +\textrm{h.c.}) \right] $\\$+\boldsymbol{\delta g^Z_{\nu L}}\frac{\hf^2}{v^2} \left[Z^\mu \bar \nu_L\gamma_\mu \nu_L+\frac{c_{\theta_W}}{\sqrt{2}}(W^{+\mu}\bar{\nu_L}\gamma_\mu e_L +\textrm{h.c.}) \right]\, $\\ $\Delta{\cal L}_{qq}^V=\boldsymbol{\delta g^Z_{uR}}\frac{\hf^2}{v^2} Z^\mu \bar u_R\gamma_\mu u_R+
\boldsymbol{\delta g^Z_{dR}}\frac{\hf^2}{v^2} Z^\mu \bar d_R\gamma_\mu d_R+\boldsymbol{\delta g^Z_{dL}}\frac{\hf^2}{v^2} \left[Z^\mu \bar d_L\gamma_\mu d_L-\frac{c_{\theta_W}}{\sqrt{2}}(W^{+\mu}\bar{u}_L\gamma_\mu d_L +\textrm{h.c.}) \right]$\\
$+\boldsymbol{\delta g^Z_{uL}}\frac{\hf^2}{v^2} \left[Z^\mu \bar u_L\gamma_\mu u_L+\frac{c_{\theta_W}}{\sqrt{2}}(W^{+\mu}\bar{u_L}\gamma_\mu d_L +\textrm{h.c.}) \right]\,$ .\\\hline TGC-Primaries\\  \hline\\
$\!\Delta{\cal L}_{\!g_1^Z}\!= \!\boldsymbol{\delta g_1^Z}\! \Bigg[  ig c_{\theta_W}\!\Big(\!Z^\mu (W^{+ \nu} W^{-}_{\mu\nu}\!-\!\textrm{h.c.}\!)\!+\!Z^{\mu\nu}W^+_\mu W^-_\nu\!\Big) -2gc_{\theta_W}^2\!\frac{\h}{v}
\Bigg(
 W^-_\mu J^\mu_- \!+\!\textrm{h.c.}
\!+\!\frac{c_{2\theta_W}}{\ctw^3} Z_\mu J^\mu_Z 
\!+\frac{2\stw^2}{\ctw}  Z_\mu J_{em}^\mu
\Bigg)$\\$
\,\times\!\!\left(1+\frac{\h}{2v}\right)+\frac{e^2v}{2\ctw^2}\h Z_\mu Z^\mu
+g^2c_{\theta_W}^2v\, \Delta_h
 \!-\!
g^2c_{\theta_W}^2\!\Big(\!W_\mu^+ W^{-\mu}+
 \frac{c_{2\theta_W} }{2\ctw^4}Z_\mu Z^\mu\!\Big) \!\!\left(\frac{5\h^2}{2}+\frac{2\h^3}{v}+\frac{\h^4}{2v^2}\!\right)\!\!$\\
$\Delta{\cal L}_{\kappa_\gamma}=\frac{\boldsymbol{\delta \kappa_{\gamma}}}{v^2}\Big[ie \hf^2(A_{\mu\nu}- t_{\theta_W}Z_{\mu\nu})W^{+\mu} W^{-\nu}+Z_\nu \partial_\mu \hf^2 (t_{\theta_W}A^{\mu \nu}-t_{\theta_W}^2 Z^{\mu\nu})$\\$+\frac{(\hf^2-v^2)}{2}
\times\Big(\ttw  Z_{\mu\nu}A^{\mu\nu}+\frac{c_{2 \theta_W}}{2 \ctw^2} Z_{\mu\nu}Z^{\mu\nu}+W^+_{\mu\nu}W^{-\mu\nu}\Big)\Big]\,$\\
$\Delta {\cal L}_{\lambda_\gamma}=\frac{i\boldsymbol{\lambda_\gamma}}{m^2_W}\left[ (eA^{\mu\nu}+g\ctw Z^{\mu\nu})
  W^{- \rho}_{\nu}  W^{+}_{\rho \mu}\right]$ \\ \hline 
\end{tabular}
\caption{\small We list the 18 most important BSM Primary effects. Here $\Delta_h$ contains terms with at least two Higgs bosons (see Ref.~\cite{Gupta} for the full expression). CP and minimal flavor violating BSM Primary effects can be found in Ref.~\cite{Gupta}.  The rest of the dimension 6  deformations comprise four fermion deformations and a gluon self interaction term.}
\end{table}

\noindent
(2)  A linear combination of the $W/Z$-couplings to fermions and the Higgs field does not have any measurable impact on $W/Z$ decays or $W/Z$ mass measurements. This is  the  linear combination that is obtained by performing the shift,
\be
{s^2_{\theta_W}}\to s^2_{\theta_W}(1+ 2\boldsymbol{\delta g_1^Z}c ^2_{\theta_W}\hf^2/v^2),
\ee
keeping $e$ constant,  only in the part of the SM Lagrangian involving $W/Z$-couplings to fermions and the Higgs field $\hat{h}$,
\begin{align}
\label{fermioncouplingscombo} 
-\boldsymbol{\delta g_1^Z} c^2_{\theta_W}\frac{\hf^2}{v^2}\Bigg[ 
 \frac{g^2\hf^2}{2}\Big(W_\mu^+ W^{-\mu}+
 \frac{c_{2\theta_W} }{2\ctw^4}Z_\mu Z^\mu\Big)
 + g (W^-_\mu J^\mu_- +\textrm{h.c.})
+\frac{gc_{2\theta_W}}{\ctw^3} Z_\mu J^\mu_Z +2 e\ttw  Z_\mu J_{em}^\mu
\Bigg]\,.
\end{align}
Clearly, for $\hat{h}=v$, this amounts to just a redefinition of $\stw$ only in the fermionic and Higgs sector and it thus has no measurable impact in processes related to $W/Z$ interactions of fermions and Higgs field. This deformation can, however, be measured by measuring the triple and quartic gauge couplings. This is because the shift ${s^2_{\theta_W}}\to s^2_{\theta_W}(1+ 2\boldsymbol{\delta g_1^Z}c ^2_{\theta_W})$ is equivalent to the opposite shift ${s^2_{\theta_W}}\to s^2_{\theta_W}(1-2\boldsymbol{\delta g_1^Z}c ^2_{\theta_W})$ only in the pure gauge sector which gives rise to the $g_1^Z$ TGC and other QGCs in a correlated way:
\be\label{TGCQGC}
\boldsymbol{\delta g_1^Z}=\frac{\delta g^{ZWW}}{ g^{ZWW}_{SM}}=
\frac{\delta g^{WWWW}}{2 \ctw^2 g^{WWWW}_{SM}}=\frac{\delta g^{ZZWW}}{2 g^{ZZWW}_{SM}}
=\frac{\delta g^{\gamma ZWW}}{ g^{\gamma ZWW}_{SM}}\, .
\ee
The expression in \eq{fermioncouplingscombo}, however, contributes to $\boldsymbol{\delta g^h_{VV}}$. The final expression for the $g_1^Z$ primary effect after making the redefinition $\boldsymbol{\delta g^h_{VV}}\to \boldsymbol{\delta g^h_{VV}}+g^2v\boldsymbol{\delta g_1^Z}\ctw^2$ to remove this projection, is given in Table 1.

To summarise the deviations from gauge coupling universality can be paremetrized by 7 deformations of $Z$-boson couplings that can be constrained by $Z$-pole measurements and one deformation that can be constrained by the $g_1^Z$ TGC. This parametrization is a generalisation of the $S$, $T$ parametrisation~\cite{Peskin}. Our parametrization has the advantage of having a one to one correspondence with all the $Z$ decays as  we have eliminated all propagator corrections by the use of equations of motion.

\section{TGCs and other remaining deformations }
There are two BSM primary effects corresponding to the two TGCs $\kappa_\gamma$ and $\lambda_\gamma$ shown in Table 1 (see Ref~\cite{Gupta} for a derivation of these two primary effects). Apart from the 18 CP even deformations in Table 1  there are also CP and minimal flavour violating deformations, a gluon self interaction term (see  Ref~\cite{Gupta} for the BSM Primary effects corresponding to these deformations)  and 25 four-fermion deformations (whose relationship with experiments is straightforward) taking the total number of independent dimension 6  BSM deformations 59.

\section{Expectations in explicit models}

The size of the BSM primary parameters  can be estimated using Naive Dimensional Analysis (NDA), assuming elementary gauge bosons but   allowing for the possibility of the Higgs field and fermions being composite \cite{Elias-Miro:2013mua},
 \be
\frac{\boldsymbol{\delta g^{h}_{ff}}}{Y_f}\sim
\frac{\boldsymbol{\delta g_{3h}}}{\lambda_h v}\sim
\frac{\boldsymbol{\delta g^{h}_{VV}}}{gm_W}\sim
\frac{\boldsymbol{\delta g^{\!Z,W}_{\!f\!,R}}}{g}\sim
\boldsymbol{\delta g^Z_{1}}\sim
\frac{g_*^2v^2}{\Lambda^2}\, ,
 \ee
 where $g_*$ denotes a generic  BSM coupling which can attain a maximum value of $4 \pi$ in strongly coupled theories. The BSM primaries $\boldsymbol{\delta \kappa_i}$ and  $\boldsymbol{\lambda_\gamma}$ are estimated to be of the order of $ g^2v^2/\Lambda^2$, where $g$ is the corresponding    SM gauge-coupling. Furthermore the couplings  $\boldsymbol{\delta \kappa_i}$ and  $\boldsymbol{\lambda_\gamma}$ are  suppressed by at least a  loop factor \cite{Elias-Miro:2013mua} in renormalizable weakly coupled theories.  For supersymmetric two Higgs doublet models the BSM Primary effects can be evaluated by integrating out the heavier Higgs bosons~\cite{Gupta1}. This generates the BSM Primary couplings $\boldsymbol{\delta g^{h}_{uu}}$, $\boldsymbol{\delta g^{h}_{dd}}$ and $\boldsymbol{\delta g_{3h}}$ while all other BSM primary couplings are zero (this is ignoring the loop-effects of the superpartners). The maximum allowed values of the BSM primary couplings $\boldsymbol{\delta g^{h}_{VV}}$,$\boldsymbol{\delta g^{h}_{uu}}$, $\boldsymbol{\delta g^{h}_{dd}}$ and $\boldsymbol{\delta g_{3h}}$ if no other BSM state related to EWSB is accessible at the LHC,   have been obtained for different well motivated BSM models in Refs.~\cite{Gupta2, Gupta3}.
 
 \section{Conclusions}
 In summary, the Lagrangian up to dimension-6 operators can be written as ${\cal L}={\cal L}_{\rm SM}+\Delta{\cal L}_{\rm BSM}$ with,
\begin{align}
&\!\!\!\!\!
\Delta{\cal L}_{\rm BSM}=
\Delta{\cal L}^h_{{\gamma \gamma}}+\Delta{\cal L}^h_{{Z \gamma}}+\Delta{\cal L}^h_{{GG}}
+\Delta{\cal L}^h_{ff}
+\Delta{\cal L}_{3h}
+\Delta{\cal L}^h_{VV}
+\Delta{\cal L}^V_{ee}+\Delta{\cal L}^V_{qq}\nn\\
&\!\!\!\!\!\!
+\Delta{\cal L}_{g_1^Z}+\Delta{\cal L}_{\kappa_\gamma}
+\Delta {\cal L}_{\lambda_\gamma}+\Delta {\cal L}_{3G}
+\Delta {\cal L}_{\rm 4f}+\Delta {\cal L}^V_{\rm MFV}+\Delta {\cal L}_{\rm CPV}\, ,\label{Ltot}
\end{align}
where $\Delta {\cal L}_{\rm 4f},~ \Delta {\cal L}_{\rm 3G},~ \Delta {\cal L}^V_{\rm MFV}$ and $\Delta {\cal L}_{\rm CPV}$ represent the BSM primary effects due to four-fermions, the three-gluon self-interaction term, MFV and CP violating terms, respectively. The amplitude for any process (for instance those that probe the tensor structure of Higgs couplings like $pp \to Wh/Zh$ or the vector boson fusion process $WW \to h$)  at the dimension 6 level, can now be written as a function of the BSM primary parameters using the above Lagrangian. Thus  all BSM processes  are already constrained, at the dimension 6 level,  by the BSM primaries and the constraints can be derived using the above Lagrangian.

\end{document}